\documentclass[aps,prd,twocolumn,amsfonts,amssymb,amsmath,showpacs,letterpaper,superscriptaddress,nofootinbib,altaffilletter,fontenc]{revtex4-1}

\usepackage[citecolor=blue,linkcolor=blue,colorlinks=true,urlcolor=blue]{hyperref}

\usepackage{graphicx}
\usepackage{amssymb}
\usepackage{amsmath}
\usepackage{longtable}
\usepackage{rotating}
\usepackage{color}
\usepackage{epsf}
\usepackage{fancyhdr}
\usepackage{ifthen}
\usepackage{slashed}
\usepackage{xspace}
\usepackage{tabularx}

\newcommand{\Msun}{\(M_\odot\)\xspace}

\newcommand{\ThisEvent}{GW190521\xspace}
\newcommand{\SEOBNR}{SEOBNRv4\_ROM\xspace}

\begin{document}

\title{Observing an intermediate mass black hole \ThisEvent with minimal assumptions}

\author{Marek\ Szczepa\'nczyk}
\author{Sergey Klimenko}
\author{Brendan O'Brien}
\author{Imre Bartos}
\author{V. Gayathri}
\author{Guenakh Mitselmakher}
\affiliation{University of Florida, Gainesville, FL 32611, USA}
\author{Giovanni Prodi}
\affiliation{Universit\`a di Trento, Dipartimento di Fisica, I-38123 Povo, Trento, Italy}
\affiliation{INFN, Trento Institute for Fundamental Physics and Applications, I-38123 Povo, Trento, Italy}
\author{Gabriele Vedovato}
\author{Claudia Lazzaro}
\address {Universit\`a di Padova, Dipartimento di Fisica e Astronomia, I-35131 Padova, Italy }
\address {INFN, Sezione di Padova, I-35131 Padova, Italy }
\author{Edoardo Milotti}
\affiliation{Dipartimento di Fisica, Universit\`a di Trieste, I-34127 Trieste, Italy}
\affiliation{INFN, Sezione di Trieste, I-34127 Trieste, Italy}
\author{Francesco Salemi}
\affiliation{Universit\`a di Trento, Dipartimento di Fisica, I-38123 Povo, Trento, Italy}
\author{Marco Drago}
\affiliation{Gran Sasso Science Institute (GSSI), I-67100 L'Aquila, Italy}
\affiliation{INFN, Laboratori Nazionali del Gran Sasso, I-67100 Assergi, Italy}
\author{Shubhanshu Tiwari}
\affiliation{Physik-Institut, University of Zurich, Winterthurerstrasse 190, 8057 Zurich, Switzerland}
\affiliation{Universit\`a di Trento, Dipartimento di Fisica, I-38123 Povo, Trento, Italy}
\affiliation{INFN, Trento Institute for Fundamental Physics and Applications, I-38123 Povo, Trento, Italy}

\begin{abstract}

On May 21, 2019 Advanced LIGO and Advanced Virgo detectors observed a gravitational-wave transient \ThisEvent, the heaviest binary black-hole merger detected to date with the remnant mass of 142\,\Msun that was published recently. This observation is the first strong evidence for the existence of intermediate-mass black holes.  The significance of this observation was determined by the coherent WaveBurst (cWB) - search algorithm, which identified \ThisEvent with minimal assumptions on its source model. In this paper, we demonstrate the capabilities of cWB to detect binary black holes without use of the signal templates, describe the details of the \ThisEvent detection and establish the consistency of the model-agnostic reconstruction of \ThisEvent by cWB with the theoretical waveform model of a binary black hole.

\end{abstract}

\date[\relax]{Dated: \today }

\maketitle

\section{Introduction}

The third observing run (O3) of the Advanced LIGO~\cite{TheLIGOScientific:2014jea} and Advanced Virgo~\cite{TheVirgo:2014hva} network has brought discoveries of new binary sources~\cite{GW190425,GW190412,GW190521.1-Discovery}, together with a wealth of gravitational-wave (GW) detection candidates~\cite{GraceDB-03-Public}. The \ThisEvent~\cite{GW190521.1-Discovery} signal observed during the first half of the O3 Advanced LIGO run has an estimated remnant mass of 142\,\Msun making it the most massive black hole found through GWs to date. \ThisEvent~\cite{GW190521.1-Discovery} provides strong observational evidence for the existence of intermediate mass black holes (IMBHs) that 
are usually defined as black holes with mass in the range $10^2-10^5$\,\Msun~\cite{Ebisuzaki:2001qm,Merzcua:2017}.

Following the first observation of a binary black hole (BBH) merger~\cite{GW150914}, the first two observing runs of Advanced LIGO/Virgo~\cite{GWTC1} have revealed a population of BBHs with component masses up to $50$\,\Msun and remnant black hole mass up to $85$\,\Msun. These observations are consistent with a mechanism known as pair-instability supernova (PISN)~\cite{Spera:2017fyx,Woosley:2016hmi,Giacobbo:2017qhh,Belczynski:2016jno}, which prevents the formation of heavier black holes from stellar core collapse. Stars with a helium core mass in the range $64 - 135$\,\Msun (PISN mass gap) undergo pulsational pair-instability and leave no remnant. Stars with core mass above $135$\,\Msun are thought to directly collapse to IMBHs. For \ThisEvent the estimated masses of the component black holes are $85^{+21}_{-14}$\,\Msun and $65^{+17}_{-18}$\,\Msun suggesting that the primary black hole may be well inside the PISN mass gap. Observations of such IMBH binary sources start to probe the boundaries of the PISN mass gap and may reveal the BBH formation mechanisms outside of the stellar evolution.

\begin{figure*}[hbt]
    \centering
    \includegraphics[width=\textwidth]{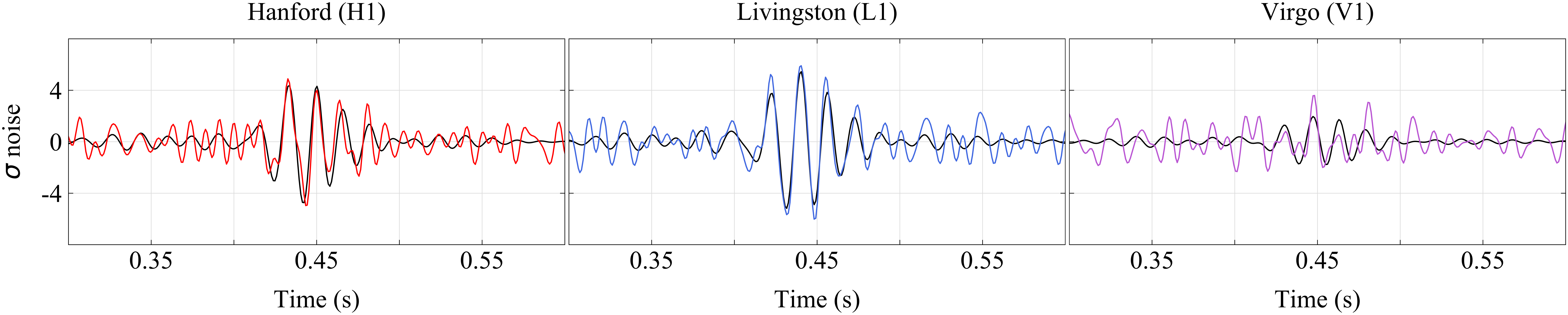}
    \caption{\ThisEvent observed by the H1 (left panel), L1 (middle panel) and V1 (right panel) detectors. Times are shown relative to May 21, 2019 at 03:02:29 UTC. Color curves: the detector time series are filtered with a 28–128 Hz band-pass filter and cleaned to remove 60\,Hz power line. Black curves: signal reconstruction by cWB. The cWB reconstruction is in agreement with the amplitudes of \ThisEvent signal inside the detector noise.
    }
    \label{fig:BLdatacWB}
\end{figure*}

Due to their high mass the IMBH binary systems are expected to merge at frequencies below 100\,Hz while detector sensitivity is limited by seismic noise below 20\,Hz. As a result, the detectors are mostly sensitive to the gravitational waves emitted at the final stages of the binary evolution, merger and ring-down. For IMBH binary systems the observed GW signal is short in duration and challenging to detect by template-based search pipelines. The excess power cWB~\cite{PhysRevD.93.042004} is a template-independent search algorithm that uses minimal assumptions on the signal model to detect GWs. It does not depend on detailed waveform features such as the higher order modes, high mass ratios, misaligned spins, eccentric orbits and possible deviations from general relativity, and it operates even in cases where the lack of reliable models pose limitations for matched filtering methods. Therefore, cWB is suitable for detecting sources where reliable templates are not readily available~\cite{2018PhRvD..97b4016C,PhysRevD.102.044035}.

First searches for gravitational waves from IMBH binaries were carried out with cWB on the data from the initial LIGO and Virgo observing runs (2005-2010)~\cite{Virgo:2012aa,aasi:2014iwa}. A search for perturbed IMBH remnants was performed on the same data with the ring-down template search~\cite{aasi:2014bqj}. Later, the IMBH searches were conducted with O1 and O2 data both by cWB and the full inspiral-merger-ringdown template pipelines~\cite{Abbott:2017iws,Abbott:2019ovz}. The IMBH binary searches carried out on the O1 and O2 data produced the most stringent upper limit of $0.2\ \mathrm{Gpc}^{-3}\mathrm{yr}^{-1}$ on the IMBH merger rates~\cite{Abbott:2017iws}. Improvements of the Advanced detector sensitivities in the O3 run~\cite{ObservingScenarios} tripled the IMBH search volume resulting in the observation of the first IMBH binary event \ThisEvent~\cite{GW190521.1-Discovery,21gImplications} 
identified by the cWB search with high confidence. Together with the first ever GW detection GW150914~\cite{GW150914} and the heaviest BBH system GW170729 in the O1 and O2 data~\cite{GWTC1}, the \ThisEvent is yet another demonstration of cWB to detect unexpected binary sources.

In this paper, we demonstrate the capability of cWB to detect \ThisEvent with high confidence without the use of templates. We show that the cWB reconstruction of \ThisEvent is in agreement with the waveforms derived from LALInference~\cite{PhysRevD.91.042003} analysis. The main results of cWB analysis are presented in the discovery~\cite{GW190521.1-Discovery} and astrophysical implication~\cite{21gImplications} papers. In Section~\ref{sec:observation} we describe the details of \ThisEvent observation with cWB, including the low-latency analysis, the \ThisEvent sky localization and describe the detection significance established by cWB. In Section~\ref{sec:waveform} we show consistency between the cWB waveform reconstruction and the best fits of the Bayesian inference waveforms from LALInference. We demonstrate and quantify that the cWB reconstruction is more consistent with waveforms from precessing BBHs rather than binaries with aligned spins. Appendices~\ref{sec:cwb},~\ref{sec:noises} and~\ref{sec:cuts} outline the cWB searches in O3 and the detection procedure.

\section{Observation}
\label{sec:observation}

\subsection{Online detection and sky localization}

The initial \ThisEvent trigger~\cite{GCN24621} was detected in low-latency by PyCBC~\cite{PyCBC} and cWB~\cite{PhysRevD.93.042004} pipelines with the FARs 1 per~8.3 years and lower than 1 per~28 years respectively. The on-line cWB FAR was estimated using one day of coincident data around the trigger time. By combining all previously accumulated background data from the low-latency analysis, the \ThisEvent FAR was estimated to be lower than 1 per 500 years.

\begin{figure}[hbt]
    \centering
    \includegraphics[width=0.45\textwidth]{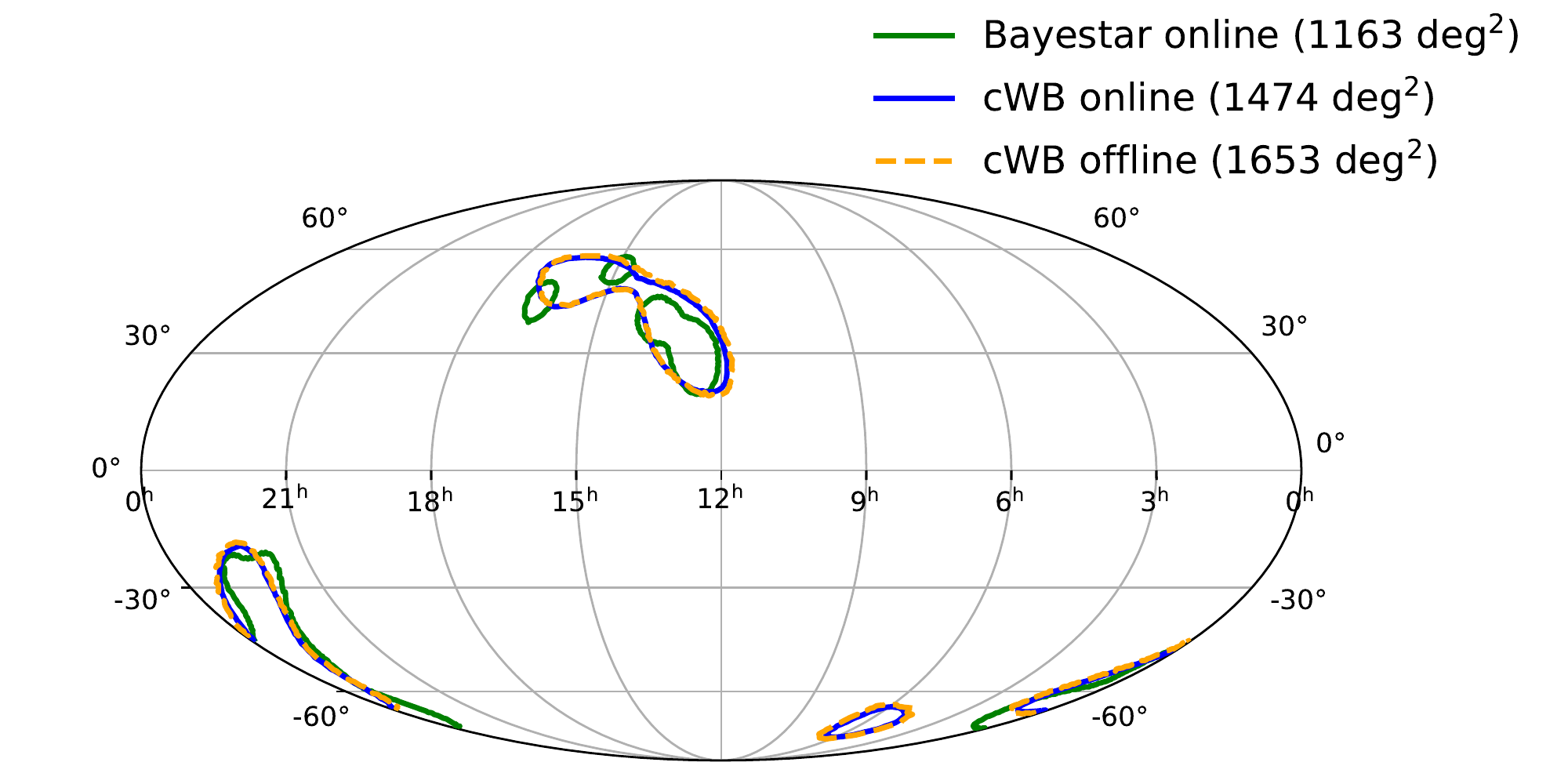}
    \caption{\ThisEvent 90\% credible areas source localization with the 3-detector LIGO-Virgo network reported by the low-latency analysis from Bayestar by using BBH templates~\cite{Bohe:2016gbl} and cWB sky localization without templates for low-latency and offline analyses.}
    \label{fig:CBCskycWB}
\end{figure}

The waveform reconstructed by cWB in low-latency analysis (see Figure~\ref{fig:BLdatacWB}) did not show a typical chirping signal expected for BBH signals. Instead, \ThisEvent was a short signal with duration of 0.1 seconds and less than 4 cycles in the frequency band 30-80\,Hz with almost symmetric shape. Such signal morphology is typical for the high mass BBH events when the system merges at low frequency where the sensitivity of the detectors is affected by seismic noise. Assuming that the binary merges at the peak signal frequency of 58\,Hz (twice the orbital frequency), the total mass of the system in the detector frame should be exceeding 300\,\Msun for equal mass BH components. The most plausible explanation for this signal is the merger of a binary BH system where the remnant is an intermediate mass black hole.

\begin{figure}[hbt]
    \centering
    \includegraphics[width=0.45\textwidth]{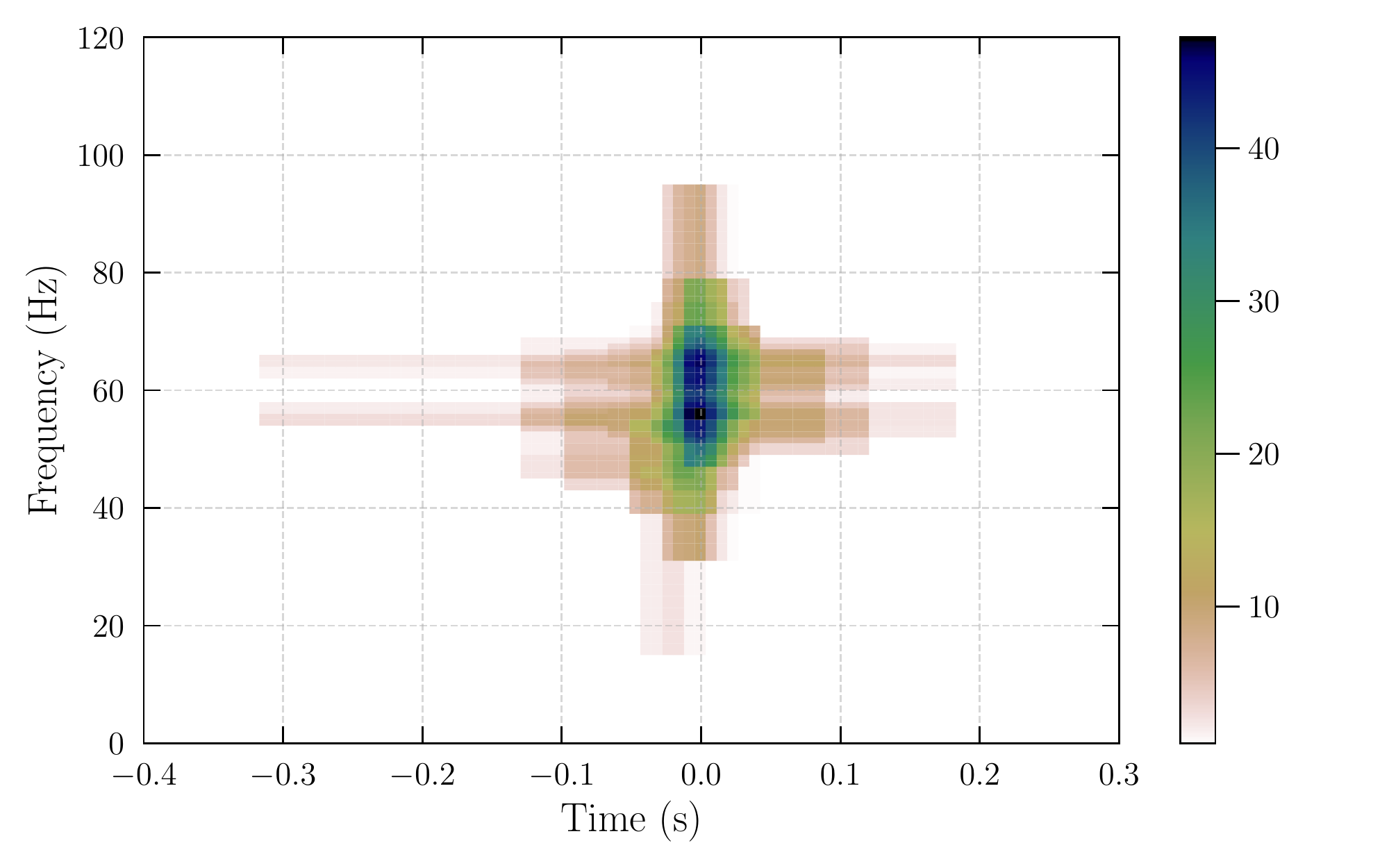}
    \caption{Time-frequency representation of \ThisEvent with a superposition of wavelet pixels selected by cWB. The pixel magnitudes are squared network signal-to-noise ratio. \ThisEvent does not show a chirping signal expected for binary systems.}
    \label{fig:scalogram}
\end{figure}

In the low-latency analysis cWB estimated the sky area to be 1474\,deg$^2$ (90\% confidence interval). It is agreement with 1163\,deg$^2$ (90\% credible interval) reported initially with the public alert~\cite{GCN24621} and computed with the Bayestar~\cite{BAYESTAR} sky localization algorithm using BBH templates~\cite{Bohe:2016gbl}. In the follow-up analysis, the sky maps were computed using calibrated data resulting in 1653\,deg$^2$ and 765\,deg$^2$~\cite{21gImplications} for cWB and Bayestar, respectively. Figure~\ref{fig:CBCskycWB} compares the initially released skymaps with those computed by cWB for the HLV detector network.

\subsection{Detection significance}

The offline cWB analysis was conducted with the  HL (LIGO Hanford and Livingston) and HLV (HL and Virgo) detector networks. Because of the difference in sensitivity between LIGO and Virgo detectors and larger non-stationary noise in Virgo~\cite{ObservingScenarios}, the HL network was used to determine the detection significance of the \ThisEvent event, while the HLV network was used for waveform reconstruction described in Section~\ref{sec:waveform}. As described in~\cite{GW190521.1-Discovery}, the data from GW detectors was conditioned prior to further analysis. Specifically for cWB analysis, we additionally mitigated the non-stationary noise coming from the anthropogenic activity (see Appendix~\ref{sec:noises}).

The detection significance of \ThisEvent was determined by time-shifting Hanford detector data with respect to the Livingston detector data to accumulate triggers of non-astrophysical origin. The time shifts were selected to be much longer (1 second or more) than the expected signal time delay between the detectors. To perform the background analysis we used 9.4~days of coincident data around the event inside the GPS time interval $[1241011102,1242485126]$. By using multiple time shifts, we accumulated an equivalent of $9,800$~years of the background data.

\begin{figure}[hbt]
\begin{center}
\includegraphics[width=\columnwidth]{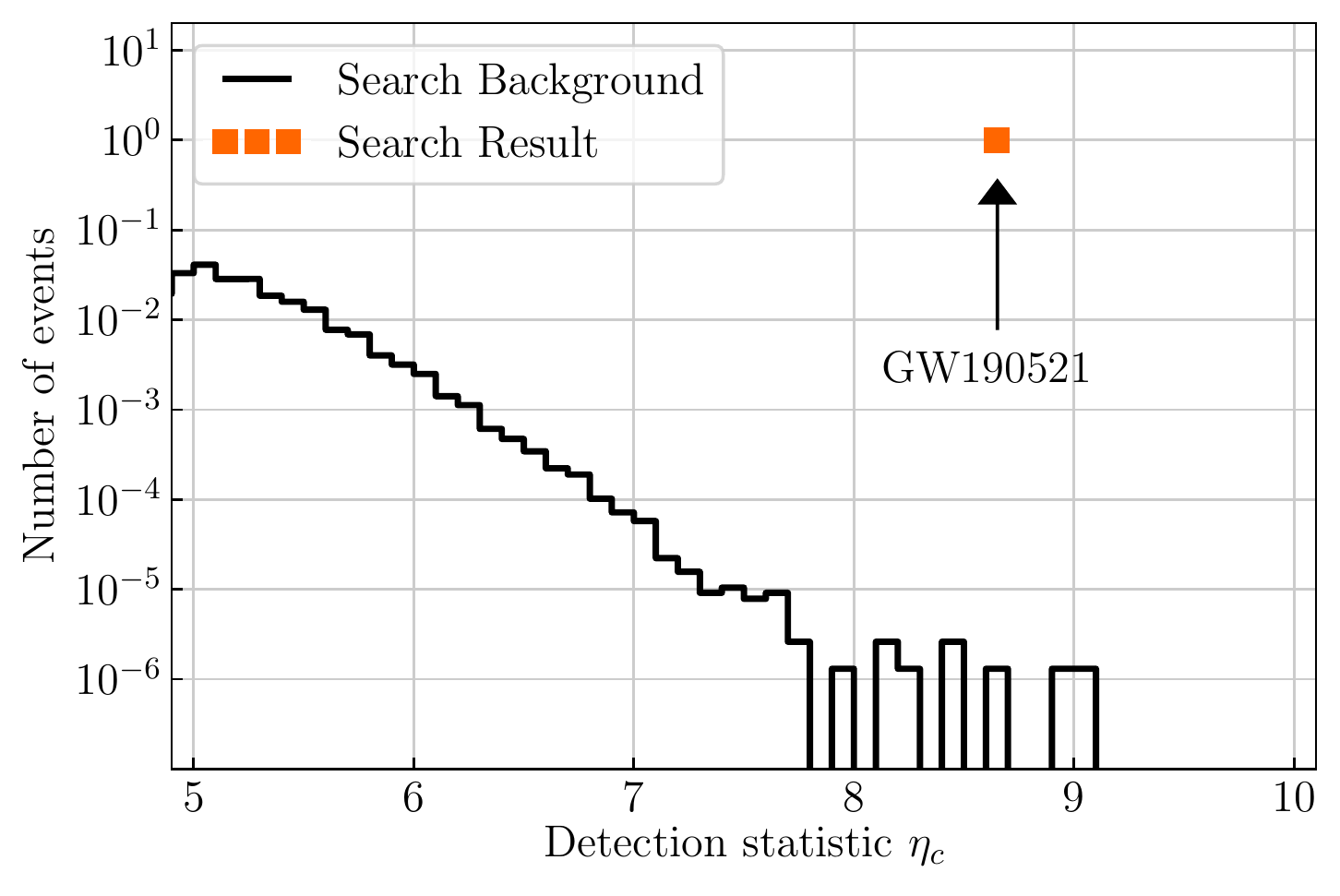}
   \caption{The search background that established the false-alarm rate of \ThisEvent to be 1 per 4900~years.}
   \label{fig:bkg}
   \end{center}
\end{figure}

Figure~\ref{fig:bkg} presents the background distribution of time-shifted data and the \ThisEvent.
Only two events have the cWB detection statistic $\eta_c$ higher than the \ThisEvent event, both consistent with random coincidences of short duration ($\approx$ 1 cycle) glitches observed in the LIGO frequency band 20–100~Hz. The amount of the background and the number of louder events results in a false-alarm rate of 1 per 4,900 years for \ThisEvent, which constitutes a confident detection.

\subsection{Search sensitivity}

The detection range of IMBH sources has been increasing over time~\cite{Virgo:2012aa,aasi:2014iwa,aasi:2014bqj,Abbott:2017iws,Abbott:2019ovz},  mainly due to improvements in the low-frequency regime of the GW detectors and the algorithms used. For the Advanced  LIGO and Advanced Virgo, assuming the NRsur7dq4 signal model~\cite{Varma:2019csw}, the detection ranges for events similar to \ThisEvent are 1.1\,Gpc, 1.2\,Gpc and 1.7\,Gpc for O1, O2 and the first half of O3, respectively~\cite{21gImplications}. At present, the total search time-volume for events like \ThisEvent is $9.1\, \mathrm{Gpc}^{3}\,\mathrm{yr}$~\cite{21gImplications}, which is the sum of the O1-O2 contribution, $3.2\, \mathrm{Gpc}^{3}\,\mathrm{yr}$ and of the nearly twice as large contribution from the first half of the O3 run: $5.9\, \mathrm{Gpc}^{3}\,\mathrm{yr}$. 

The  rate of the IMBH binary mergers was initially constrained from $1.3 \times 10^{2}\, \mathrm{Gpc}^{-3}\,\mathrm{yr}^{-1}$ using the 2005-2007 data set~\cite{Virgo:2012aa}, and later to $0.2\, \mathrm{Gpc}^{-3}\,\mathrm{yr}^{-1}$ with data from the first two LIGO-Virgo observing runs~\cite{Abbott:2017iws}. Now we estimate that the rate of events similar to \ThisEvent is $0.13_{-0.11}^{+0.3}\, \mathrm{Gpc}^{-3}\,\mathrm{yr}^{-1}$~\cite{21gImplications} that is below the previously derived upper limits.

\section{Waveform analysis}
\label{sec:waveform}

\begin{figure*}[bt]
    \centering
    \includegraphics[width=\textwidth]{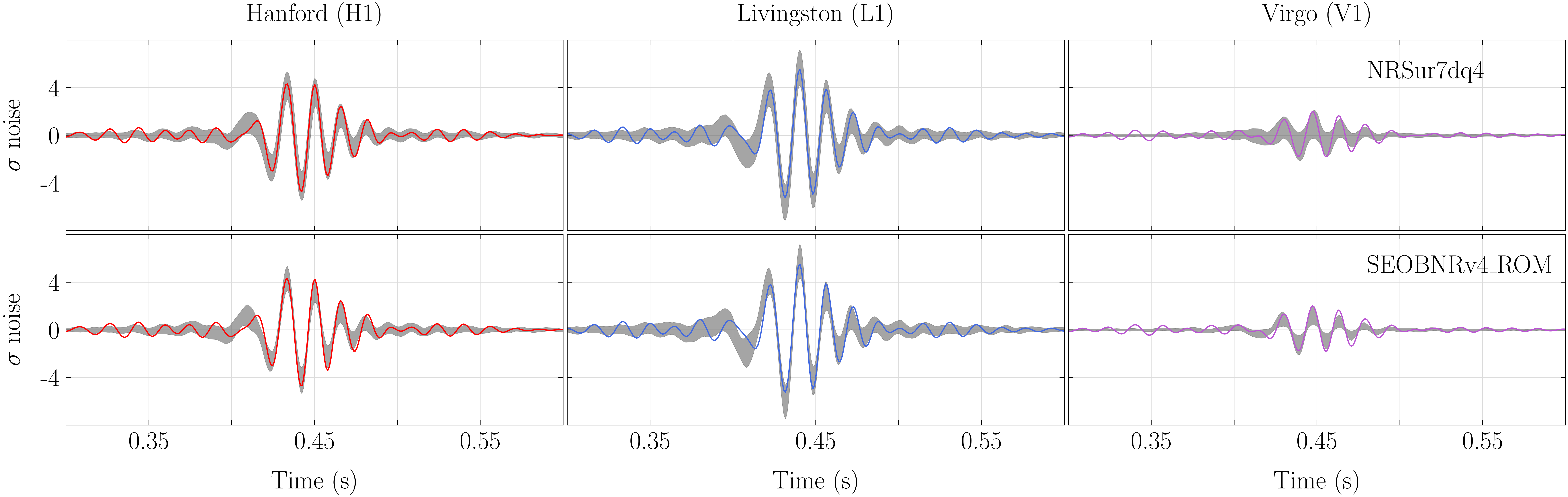}
    \caption{Visualization of the consistency of the signal-agnostic measurement by cWB with two alternative models in time domain: upper panel NRSur7dq4 model, lower panel \SEOBNR model. The colored curves are the whitened waveform for \ThisEvent as reconstructed by cWB for the LIGO Hanford (red), LIGO Livingston (blue) and Virgo (violet); they are a direct measurement and do not change from the upper to the lower panels. The shaded belts are 90\% confidence intervals per each time sample at each detector and are model dependent.
    }
    \label{fig:hoftcomparison}
\end{figure*}

In this section, we present the comparison of the \ThisEvent reconstruction by cWB with the LALInference model-based Bayesian reconstruction~\cite{PhysRevD.91.042003, Ashton:2018jfp}. The purpose of this comparison is to demonstrate if the cWB and LALInference reconstructions are consistent for different signal models.

The cWB waveforms are derived directly from the data in the wavelet (time-frequency) domain~\cite{Necula:2012zz} by selecting the excess power data samples above the average fluctuations of the detector noise. The cWB point estimate of the signal waveform in each detector is constructed as a linear combination of selected wavelets with the amplitudes defined by the constrained maximum likelihood method~\cite{PhysRevD.93.042004}. Similar to the band-pass filter shown in Figure~\ref{fig:BLdatacWB} cWB performs the time-frequency filtering of data. Such signal-agnostic reconstruction may capture significant features of the signal that may not be well described by the model approximations and help to identify limitations of the models used for Bayesian inference of \ThisEvent. On the other hand, due to the excess power threshold, the low SNR wavelet components could be excluded from the cWB analysis resulting in a partial reconstruction of the \ThisEvent signal. In addition, the non-Gaussian noise fluctuations may bias the cWB reconstruction and add systematic errors not accounted for by the Bayesian inference. Therefore, a direct comparison of the cWB reconstruction, or any other excess power algorithm, with the best matching LALInference waveforms is not straightforward and requires a more accurate statistical treatment as described below.

In this analysis we compare cWB point estimate waveform with the effective-one-body model \SEOBNR~\cite{SEOBNRV4PHM,Babak:2016tgq} and the numerical relativity surrogate model NRSur7dq4~\cite{Varma:2019csw}, which is the preferred model for the \ThisEvent event.
For correct comparison with the cWB point estimate, the simulated signals from the both models are treated exactly in the same way as the real GW signal: simulated signals are selected randomly from the LALInference posterior distribution, injected into the data (in the time interval of 4096 seconds around the \ThisEvent event) and analysed with cWB. Then we use the cWB reconstruction to produce the off-source point estimate for each of the simulated events.  These off-source simulations take into account the fluctuations of the real detector noise, cWB reconstruction errors and the variability of the  signal model in the LALInference posterior distribution.  The off-source simulations are synchronized in time and used for the construction of confidence intervals showing the expected spread of the modeled signal amplitudes as seen by cWB, which can be directly compared with the cWB point estimate for \ThisEvent. 
\begin{figure*}[bt]
    \centering
    \includegraphics[width=\textwidth]{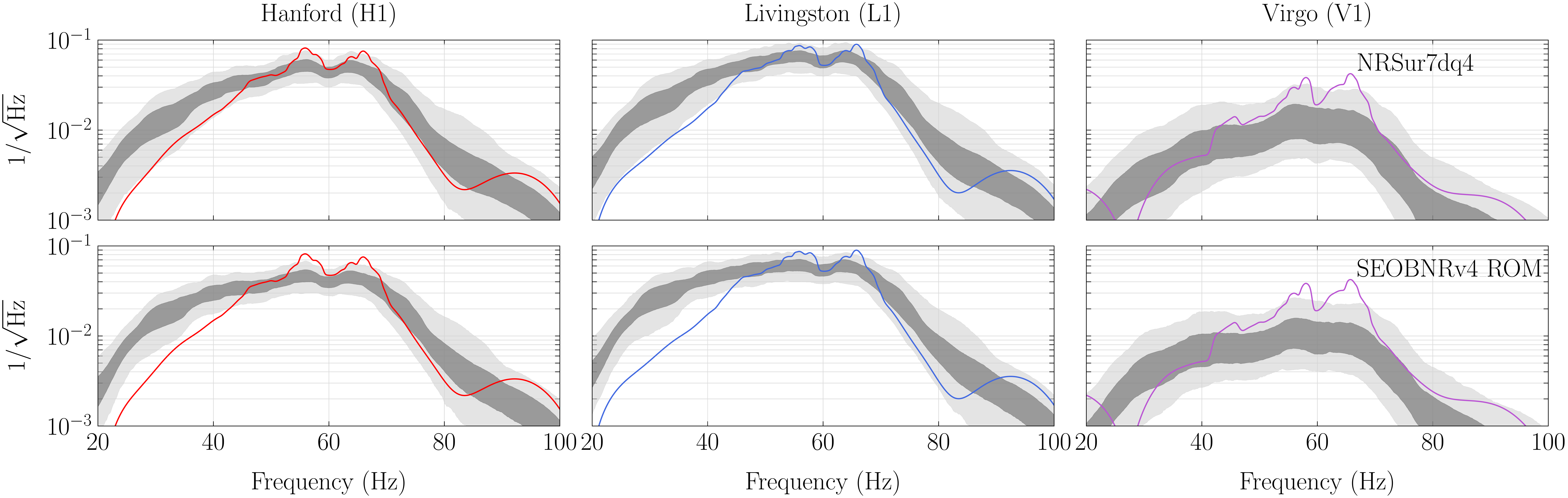}
    \caption{Visualization of the consistency of the signal-agnostic measurement by cWB with two alternative models in frequency domain: upper panel NRSur7dq4 model, lower panel \SEOBNR model. The colored curves are the whitened waveform spectra for \ThisEvent \ as reconstructed by cWB for the LIGO Hanford (red), LIGO Livingston (blue) and Virgo (violet); they are a direct measurement form the data and do not change from the upper to the lower panels. The dark shaded belts are the 50\% confidence intervals per each frequency bin showing the variation of the model spectral amplitudes. The light shaded belts are the 90\% confidence intervals.
    }
    \label{fig:FFTcomparison}
\end{figure*}
The result is shown in Figure~\ref{fig:hoftcomparison} where the  90\% confidence intervals are calculated from whitened waveforms of approximately 1400 simulated events. They display that with 90\% probability the local amplitudes of the \ThisEvent signal filtered by cWB are located within these boundaries. The cWB reconstruction of \ThisEvent comprises about 11 degrees of freedom (the effective number of independent wavelets), so confidence intervals for the contiguous time values are correlated. An approximate estimate of how frequently one should expect at least one deviation outside the belt ensuring 90\% confidence is given by $1-e^{-11\times0.1}$ or approximately 70\%. The local deviations outside of the 90\% interval visible in top plots for the  NRSur7dq4 models are negligible. However, the \SEOBNR model appears to have larger deviation at the beginning of the \ThisEvent signal. The difference between the two models is also observed in the frequency domain. Figure~\ref{fig:FFTcomparison} shows 90\% confidence intervals built from the Fourier representation of the whitened off-source reconstructions of simulated events. This analysis highlights more discrepancy between NRSur7dq4 and \SEOBNR models than in the time domain.  In particular, at lower frequency the NRSur7dq4 model shows a deficit of energy in comparison with the spin-aligned \SEOBNR model, demonstrating a better agreement with the cWB  reconstruction of \ThisEvent. Such a deficit of energy before the merger is expected for signals with orbital precession predicted by the NRSur7dq4 model or signals with large eccentricity~\cite{2020arXiv200901066C}.

\begin{figure*}[hbt]
    \centering
    \includegraphics[width=0.49\textwidth]{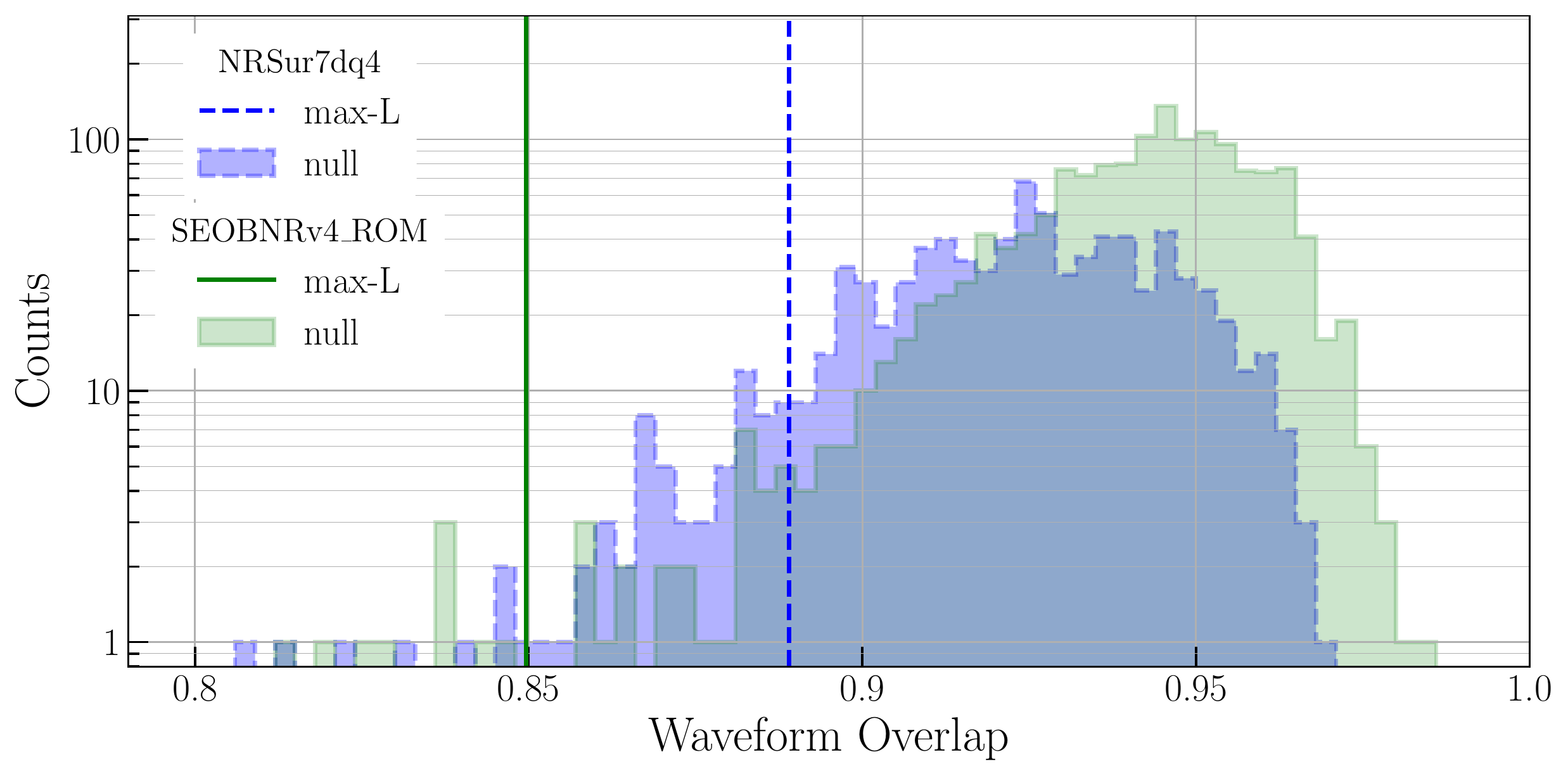}
    \includegraphics[width=0.49\textwidth]{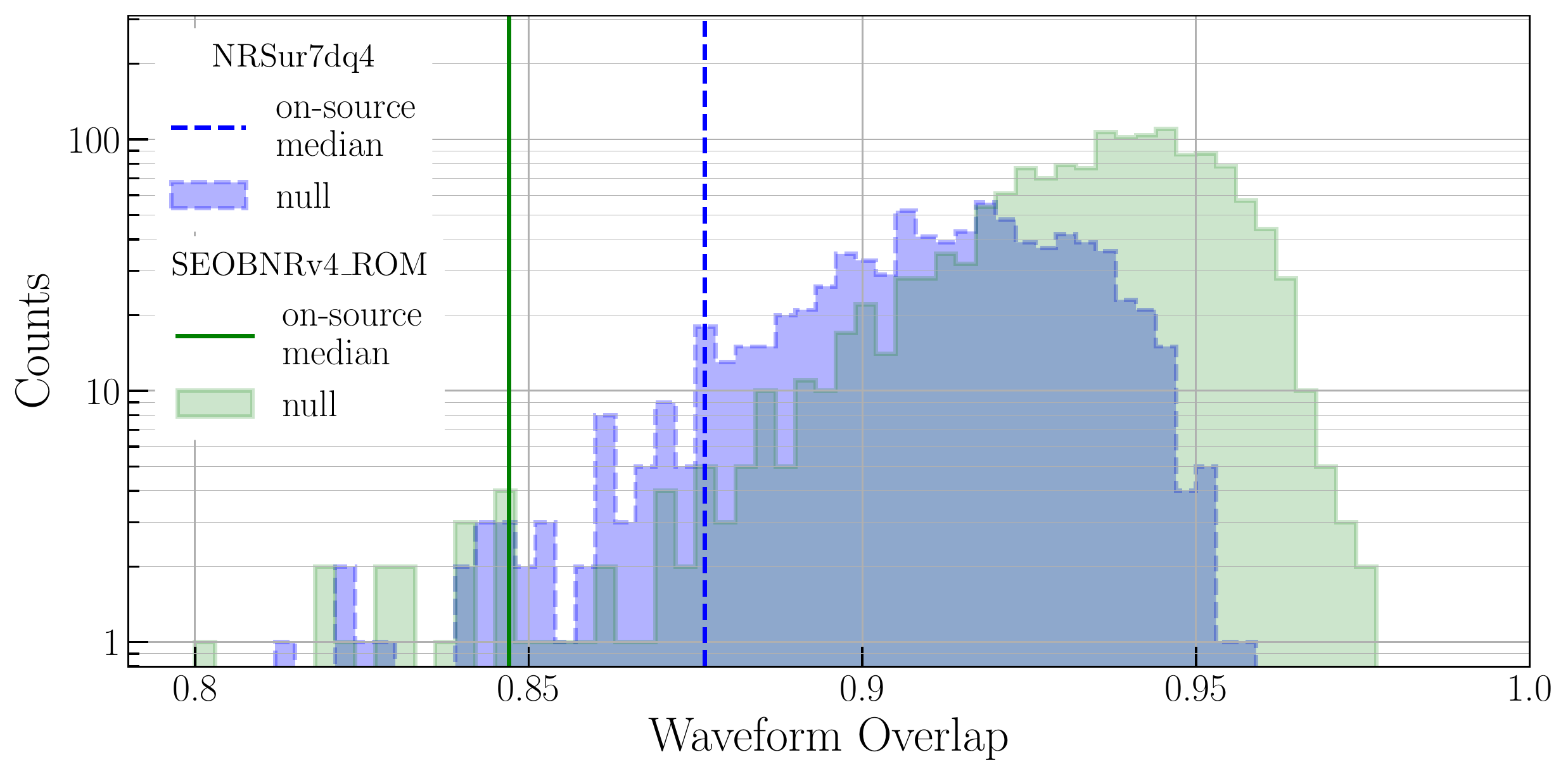}
\caption{ The quantitative comparison of whether the signal models are consistent with cWB reconstruction of \ThisEvent. The plots show the off-source (null) distributions and calculated waveform overlaps for \ThisEvent. Left panel: the max-L is assumed to be a representative waveform for \ThisEvent. Right panel: any posterior waveform is assumed to be a valid representation of \ThisEvent. In both cases, the p-values are 7.9\% and 1.0\% for NRSur7dq4 and \SEOBNR, respectively. It indicates that \SEOBNR is less consistent with \ThisEvent but its relatively large p-value does not allow to confidently exclude this model.}
\label{fig:match}
\end{figure*}

\subsection{Waveform overlap}
\label{sec:overlap}

The time and frequency confidence intervals is a convenient tool to identify local waveform differences between the cWB reconstruction of \ThisEvent and the signal model. However, they do not provide a statistical measure for disagreement between a given model and the \ThisEvent signal. To quantify the consistency between different waveforms, we use the waveform overlap, or the  match~\cite{GW190521.1-Discovery,PhysRevD.100.042003}, between the cWB point estimate  $\mathbf{w} = \{w_{k}(t)\}$ and a selected whitened waveform from a given signal model $\mathbf{h} = \{{h}_{k}(t)\}$:
\begin{equation}
 \label{FFdef}
 \mathrm{O}(\mathbf{w},\mathbf{h}) =  \frac{(\mathbf{w}|\mathbf{h}) }{\sqrt{(\mathbf{w}|\mathbf{w}) } \sqrt{(\mathbf{h}|\mathbf{h}) } } \,.
\end{equation}
The scalar product $(.|.)$ is defined in the time domain as  
\begin{equation*}
\label{sprod}
(\mathbf{w}|{\mathbf{h}}) = \sum_k \int_{t_{1}}^{t_{2}} {w}_k(t) {h}_k(t) dt,
\end{equation*}
where the index $k$ running over the detectors and $[t_1, t_2]$ is the time range of the reconstructed event. First, we calculate the distribution of $\mathrm{O}_{0}(\mathbf{w_i},\mathbf{h_i})$ for the off-source injections, where the waveform $\mathbf{h_i}$ is drawn randomly from the posterior distribution and the $\mathbf{w_i}$ is the cWB point estimate of $\mathbf{h_i}$. The $\mathrm{O}_{0}$ gives us the null distribution expected for a given signal model accounting for systematic uncertainties due to non-Gaussian detector noise and the cWB reconstruction. Second, we calculate the overlap $\mathrm{O}(\mathbf{W},\mathbf{h_p})$ where $\mathbf{W}$ is the on-source cWB reconstruction of \ThisEvent and $\mathbf{h_p}$ is a proxy of the true \ThisEvent signal selected from the posterior distribution. If the model accurately describes the \ThisEvent event, we expect that the overlap $\mathrm{O}(\mathbf{W},\mathbf{h_p})$ falls within the null distribution, which can be characterized with the p-value - the fraction of entries in the null distribution with the overlap below $\mathrm{O}(\mathbf{W},\mathbf{h_p})$. Left panel in Figure~\ref{fig:match} shows the null distribution and the overlap with the maximum likelihood (max-L) template, used as the proxy $\mathbf{h_p}$ for the \ThisEvent signal, evaluated for both signal models. The max-L overlaps are 0.89 and 0.85, while the p-values are 7.9\% and 1.0\% for the NRSur7dq4 and \SEOBNR models, respectively. In agreement with Figures~\ref{fig:hoftcomparison} and~\ref{fig:FFTcomparison},  this figure illustrates that the latter signal model is less consistent with the \ThisEvent event. However, the relatively large p-value of 1\% does not allow us confident exclusion of the \SEOBNR as a model for \ThisEvent.

The assumption of the max-L waveform being the best representation of \ThisEvent is viable but not exclusive. The maximum posterior waveform could also be used instead of the max-L. In addition to this ambiguity, a selection of any specific proxy waveform $\mathbf{h_p}$ introduces a bias in the calculation of the p-value. To eliminate this bias, for each off-source injection $\mathbf{h_i}$ we should run the LALinference parameter estimation to find the corresponding proxy waveform, which should be used instead of $\mathbf{h_i}$ for the calculation of the null distribution. This is not practically feasible due to the high computational cost of the LALinference analysis. Instead, we modify the statistical procedure in order to reduce the selection bias. In general, for a correct model, any posterior waveform could be a fair proxy $\mathbf{h_p}$ for the \ThisEvent signal. Therefore, we calculate the on-source overlap distribution $\mathrm{O}(\mathbf{W},\mathbf{h_i})$ by randomly drawing $\mathbf{h_i}$ from the posterior distribution. Instead of selecting a specific $\mathrm{O}(\mathbf{W},\mathbf{h_p})$ overlap, we calculate the median of the on-source distribution. Similarly, a set of distributions $\mathrm{O}(\mathbf{w_j},\mathbf{h_i})$ can be calculated for any cWB point estimate $\mathbf{w_j}$ from the off-source injections. By calculating the median of each $\mathrm{O}(\mathbf{w_j},\mathbf{h_i})$ distribution we can build a new null distribution for the median of $\mathrm{O}(\mathbf{W},\mathbf{h_i})$. The right panel of Figure~\ref{fig:match} shows these null distributions and the on-source median overlap values of 0.88 and 0.85 for the NRSur7dq4 and \SEOBNR models, respectively. The corresponding p-values are 7.9\% and 1.0\%. These numbers are similar to the results obtained with the analysis using the max-L waveform. Nevertheless, the comparison of the entire posterior distribution with the corresponding off-source null distributions provide a more robust test of the \ThisEvent models.

\section{Summary}  

The cWB searches for GW transients have demonstrated confident detection of  binary black holes, including the first gravitational-wave signal GW150914 detected in the O1 run~\cite{GW150914}, the heaviest BBH system GW170729 detected in the O2 run together with the other detection from the first two observing runs of Advanced LIGO and Advanced Virgo~\cite{GWTC1}. The observation of \ThisEvent, the first confident detection of an IMBH by cWB, is yet another demonstration of its capabilities to discover new GW sources. 
The cWB detection in low-latency provided a robust reconstruction of \ThisEvent including a sky map that was in agreement with the template based searches. The offline cWB analysis provided a confident detection of \ThisEvent with the false-alarm rate of 1 per 4,900 years. The estimated rate of events similar to \ThisEvent ($0.13_{-0.11}^{+0.3}\, \mathrm{Gpc}^{-3}\,\mathrm{yr}^{-1}$~\cite{21gImplications}) is below the previously derived upper limits. The waveform analysis presented in the paper demonstrates the agreement between cWB and LALInference reconstructions for different signal models. The consistency between signal models and cWB reconstruction is described by p-values, which are 7.9\% and 1.0\% for the NRSur7dq4 and \SEOBNR models respectively. While the p-value of NRSur7dq4 shows that this model is consistent with the cWB reconstruction of \ThisEvent, the relatively large p-value for \SEOBNR signal model does not allow us to confidently exclude this model.

The anticipated sensitivity improvements of the GW instruments at low-frequency~\cite{ObservingScenarios} can lead to detections of more IMBH sources. These sources could be formed dynamically and may have waveforms not covered by existing templates. The template-independent search has a potential to play an important role in the detection of such signals and help us to explore the physical properties of IMBH sources in a wide range of masses.

\section{Acknowledgments}

This research has made use of data, software and/or web tools obtained from the Gravitational Wave Open Science Center, a service of LIGO Laboratory, the LIGO Scientific Collaboration and the Virgo Collaboration. The work by S.~K. was supported by NSF Grant No. PHY 1806165. We gratefully acknowledge the support of LIGO and Virgo for provision of computational resources. IB acknowledges support by the National Science Foundation under grant No. PHY 1911796, the Alfred P. Sloan Foundation and by the University of Florida.

\clearpage
\appendix

\section{Coherent WaveBurst}
\label{sec:cwb}

Coherent WaveBurst (cWB) is a search algorithm for detection and reconstruction of GW transient signals operating without a specific waveform model. cWB searches for a coincident signal power in  multiple detectors for signals with the duration up to a few seconds. The analysis is performed in the wavelet domain~\cite{Necula:2012zz} on data normalized (whitened) by the amplitude spectral density of the detector noise. The cWB selects wavelets with amplitudes above the fluctuations of the detector noise and groups them into clusters. For clusters correlated in multiple detectors, cWB identifies coherent events and reconstructs the source sky location and signal waveforms with the constrained maximum likelihood method~\cite{PhysRevD.93.042004}.

The cWB detection statistic is based on the coherent network energy $E_{\rm c}$ obtained by cross-correlating data in different detectors; $\sqrt{E_{\rm c}}$ is the estimator of the coherent network SNR~\cite{PhysRevD.93.042004}.  The agreement between GW data and cWB reconstruction is characterized by a chi-squared statistic $\chi^2 = E_{\rm n}/N_{\rm df}$, where $E_{\rm n}$ is the residual  energy estimated after the reconstructed waveforms are subtracted from the whitened data, and $N_{\rm df}$ is the number of independent wavelet amplitudes describing the event. \ThisEvent is identified in LIGO data with $\chi^2 = 0.68$, showing no evidence for residual energy inconsistent with Gaussian noise. The cWB detection statistic is $\eta_{\rm c}=\sqrt{E_{\rm c}/\max(\chi^2,1)}$, where the $\chi^2$ correction improves the robustness of the pipeline against non-Gaussian detector noise. Events with $\eta_{\rm c}>5$ are stored by the pipeline for further processing. 

To improve the robustness of the algorithm against non-stationary detector noise (glitches) and reduce the rate of false alarms, cWB uses signal-independent vetoes. The primary veto cuts are on the network correlation coefficient $c_{\rm c} = E_{\rm c}/(E_{\rm c}+E_{\rm n})$ and the $\chi^2$. For a GW signal the expected values of both statistics are close to unity and the candidate events with $c_{\rm c} < 0.7$ and $\chi^2>2.5$ are rejected as potential glitches.

\section{Mitigation of non-stationary detector noise}
\label{sec:noises}

\begin{figure}[ht]
\begin{center}
  \includegraphics[width=\columnwidth]{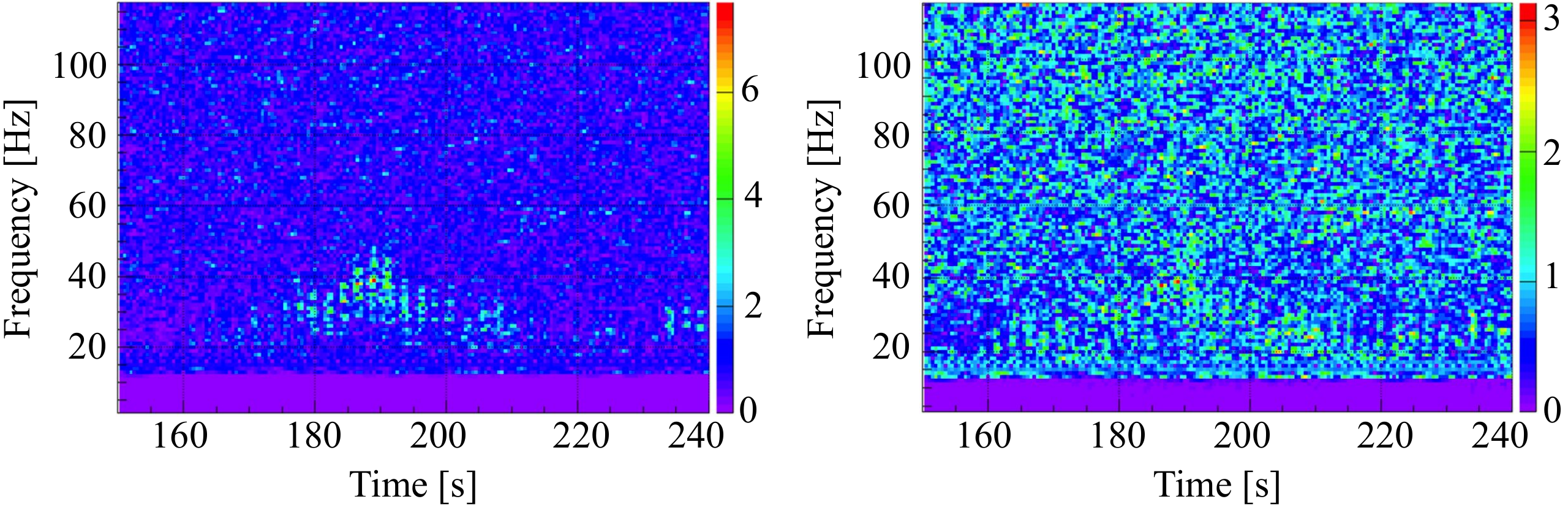}
   \caption{Time-frequency WDM wavelet decomposition of a segment of LLO data: left (right) - before (after) application of the LPE filter. The excess power due to anthropogenic ground motion is reduced.}
   \label{fig:signalFFT}
\end{center}
\end{figure}

The analysis of the early O3 data revealed several dominant sources of non-stationary noise (glitches). The dominant sources of this noise are the low frequency glitches below 50\,Hz due to anthropogenic ground motion and the short duration (blip) glitches in the frequency band $50-300$\,Hz~\cite{Abbott:2017ocd,Cabero:2019orq}. The anthropogenic ground motion in LIGO Livingston (LLO) produced the scattered light noise below 50\,Hz~\cite{GW190521.1-Discovery} appearing as a periodic sequence of short duration transients. A similar type of noise is also observed in the LIGO Hanford (LHO) detector, but at a significantly lower rate. Left panel of Figure~\ref{fig:signalFFT} shows the example of scatter light glitches in the time-frequency map of the LLO data segment. This noise is highly periodic with the two characteristic times of 2.35\,s and 0.235\,s. It can be corrected with the linear predictor error filter (LPE)~\cite{Tiwari:2015a} as shown in the right panel of Figure~\ref{fig:signalFFT}. The correction of the wavelet data was performed in the frequency band $16-48$\,Hz and does not affect the detection and reconstruction of signals at higher frequency. No scattered light glitches were observed at the time of \ThisEvent, which has the peak frequency of 58\,Hz. The application of the LPE filter significantly reduces the rate of the scattered light glitches and improves the detection of IMBH binary signals expected at low frequencies.

\begin{figure}[ht]
\begin{center}
  \includegraphics[width=\columnwidth]{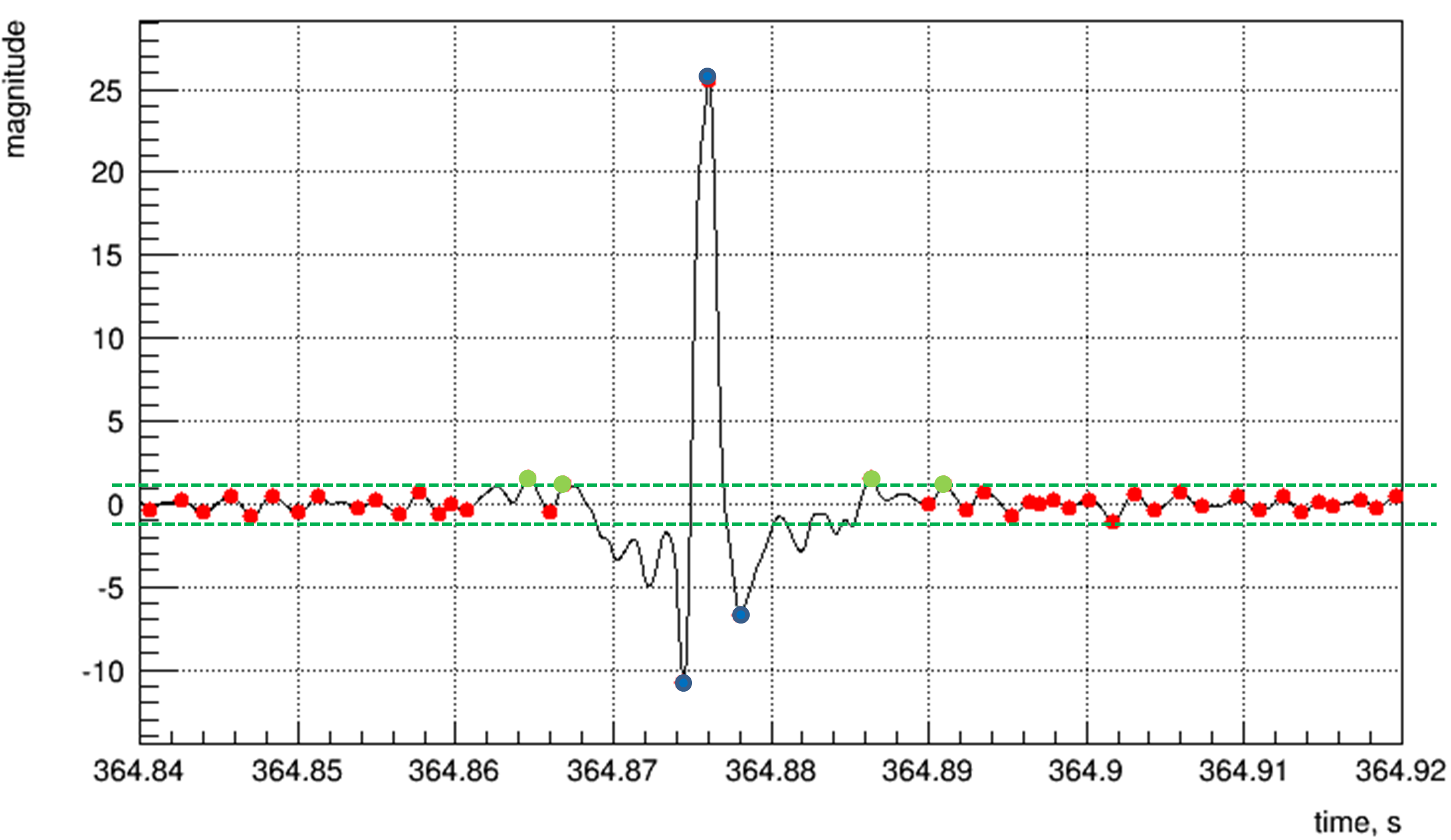}
   \caption{An example of a blip glitch. The characteristic waveforms of blip glitches allow us to distinguish them from gravitational waves and other sources of noise as described in the text.}
   \label{fig:blip}
\end{center}
\end{figure}

\begin{table*}[hbt]
\begin{tabular}{c|cccc|cccc}
\hline
\hline
\textbf{Case} & \textbf{1} & \textbf{2} & \textbf{3} & \textbf{4} & \textbf{5} & \textbf{6} & \textbf{7} & \textbf{8} \\
\hline
\textbf{IMBH configuration} & IMBH,BBH & - & IMBH & - & IMBH & BBH & BBH & - \\
\textbf{BBH configuration} & - & IMBH,BBH & - & BBH & BBH & IMBH & - & IMBH \\
\hline
\textbf{False-alarm rate} & IMBH  & BBH & IMBH  & BBH & \multicolumn{4}{c}{min(BBH,IMBH)$\times 2$}\\
\hline
\hline
\end{tabular}
\caption{Possible detection cases and corresponding trial factors. For example, in case 5, an event is identified in both search configurations but the peak frequencies are reconstructed in different bands, so the smaller false-alarm rate from the two configurations multiplied by a trial factor of 2 defines the event's significance. The calculation of \ThisEvent significance is given by case 3.}
\label{tab:searches}
\end{table*}

The short noise transients are primarily blip glitches present in all observing runs and their origin is yet unknown~\cite{Abbott:2017ocd,Cabero:2019orq}. The properties of blip glitches have been extensively studied in the previous observing runs~\cite{Abbott:2016ctn} and to remove them we apply a similar method used in the previous cWB searches~\cite{GWTC1,Abbott:2019ovz}. The waveforms are typically time-symmetric, with less than one cycle, and without a clear frequency evolution. The duration is usually of an order of $O(10)$\,ms and bandwidth $O(100)$\,Hz. Figure~\ref{fig:blip} shows an example of a blip glitch. The characteristic waveform of blip glitches allows us for their simple classification as follows.   The energy of the data samples with the largest magnitude ($A_m$) and two peak amplitudes around it (blue dots) is denoted as $E_0$. The remaining samples with the amplitudes outside of the thresholds $\pm A_m$/7.6 (green dots) have the energy denoted as $E_1$. The ratio $Q = E_1 / E_0$ is calculated for each detector used to reconstruct an event and a smallest value of $Q$ out of all detectors may indicate the presence of a blip glitch. Events with $Q < 0.1$ are removed from the further analysis. \ThisEvent with a $Q$ value of 0.55 passed this criteria.

\section{Search configurations}
\label{sec:cuts}

A generic search for binary systems covers a large parameter space. It is not possible to design a search optimized to all binaries because the frequency content of their GW signals can vary significantly. In general, GW signals have the peak frequency $f_{peak}\propto 1/M_{tot}$, where $M_{tot}$ is the binary total mass in the detector frame. Lighter binaries with $M_{tot}\approx2$\,\Msun  merge at high frequency and their GW signal may be observable in the detector data for hundreds of seconds. Heavier systems with $M_{tot}>100$\,\Msun merge at lower frequencies and their detected signals are less than a second long. Therefore, a generic search for binary systems is split into the two search configurations. One configuration focuses on the IMBH binaries ($f_{peak}<80$\,Hz), while the other configuration is used for detection of stellar mass BBHs ($f_{peak}>80$\,Hz). Because the detector noise and glitches vary significantly in these two frequency bands, the searches require different selection cuts.

The two configurations are not disjoint and trial factors need to be considered to calculate events significance and they are determined by the reconstructed peak frequencies. Table~\ref{tab:searches} summarizes all cases when a binary system is detected either by the BBH or IMBH search configuration. For example, if the event is detected in both search configurations with the peak frequencies above 80\,Hz, then the BBH search is selected and the significance of the event is not penalized by the trial factor. In contrast, if the peak frequencies of detected signals are reconstructed in different frequency bands, then the smaller FAR is used for calculation of the event's significance and it is multiplied by a trial factor of 2.

\newpage
\bibliography{detections}

\end{document}